\documentclass[sigconf]{acmart}

\usepackage{booktabs} 

\usepackage{balance}  
\usepackage{graphics} 
\usepackage{multirow}
\usepackage[justification=centering]{caption}
\usepackage{xspace}
\usepackage{calc}
\usepackage{flushend}

\renewcommand{\textrightarrow}{$\rightarrow$}

\newcommand{\FC}{{FreeGuard}}
\newcommand{\FG}{{FreeGuard}}

\newcommand{\DL}{\texttt{DLmalloc}}

\newcommand{\specialcell}[2][c]{%
  \begin{tabular}[#1]{@{}c@{}}#2\end{tabular}}

\renewcommand{\paragraph}[1]{\vspace{0.1in}\noindent{\bf{#1}.}}

\begin{document}

\title{\FG{}: A Faster Secure Heap Allocator}

\author{Sam Silvestro}
\affiliation{%
  \institution{University of Texas at San Antonio}
}
\email{Sam.Silvestro@utsa.edu}

\author{Hongyu Liu}
\affiliation{%
  \institution{University of Texas at San Antonio}
}
\email{liuhyscc@gmail.com}

\author{Corey Crosser}
\affiliation{%
  \institution{United States Military Academy}
}
\email{Corey.Crosser@usma.edu}

\author{Zhiqiang Lin}
\affiliation{%
  \institution{University of Texas at Dallas}
}
\email{zhiqiang.lin@utdallas.edu}

\author{Tongping Liu}
\affiliation{%
  \institution{University of Texas at San Antonio}
}
\email{Tongping.Liu@utsa.edu}

\renewcommand{\shortauthors}{S. Silvestro et al.}

\begin{abstract}
In spite of years of improvements to software security, heap-related attacks still remain a severe threat. One reason is that many existing memory allocators fall short in a variety of aspects. For instance, performance-oriented allocators are designed with very limited countermeasures against attacks, but secure allocators generally suffer from significant performance overhead, e.g., running up to $10\times$ slower. 
This paper, therefore, introduces \FG{}, a secure memory allocator that prevents or reduces a wide range of heap-related attacks, such as heap overflows, heap over-reads, use-after-frees, as well as double and invalid frees. \FG{} has similar performance to the default Linux allocator, with less than 2\% overhead on average, but provides significant improvement to security guarantees. 
\FG{} also addresses multiple implementation issues of existing secure allocators, such as the issue of scalability. 
Experimental results demonstrate that \FG{} is very effective in defending against a variety of heap-related attacks.

\end{abstract}


\maketitle
\section{Introduction}
\label{sec:intro}

C/C++ programs (e.g., web browsers, network servers) often require dynamically managed heap memory. However, it is very challenging to guarantee heap security. Over the past decades, a wide range of heap-related vulnerabilities -- such as heap over-reads, heap overflows, use-after-frees, invalid-frees, and double-frees -- have been discovered and exploited for attacks, including denial-of-service, information leakage, and control flow hijacking~\cite{Szekeres:2013:SEW:2497621.2498101}.
Currently, heap vulnerabilities continue to emerge. For instance, as shown in Table~\ref{tbl:comparison1}, a significant number were still observed within the past three months. It is very unlikely that heap vulnerabilities will disappear in the near future, without significant advancement in detection techniques. Thus, efficient and effective techniques are still required to defend against these vulnerabilities.

\begin{table}[h]
  \footnotesize
  \centering
  \begin{tabular}{ l | r } 
  {Vulnerabilities} & \specialcell{Occurrences (\#)}\\
   \hline
   Heap Over-reads & 54 \\
   Heap Overflows & 66 \\
   Use-after-frees & 5 \\
   Invalid-frees & 2 \\
   Double-frees & 2 \\
   \hline
   \end{tabular}
  \caption{
  	Number of heap vulnerabilities in the past \\three months (collected on 08/26/2017 from NVD~\cite{nvd}).   	  \label{tbl:comparison1}
  	}
  	\vspace{-0.2in}
\end{table}

One method used to secure the program heap is to add defenses within the memory allocator~\cite{DieHarder}, which can be combined with other security mechanisms, such as non-executable segments and address space layout randomization (ASLR). However, existing allocators are either insecure or inefficient. 
In particular, existing memory allocators can be classified into two types, based on their implementation mechanisms. 

One type belongs to bump-pointer or sequential allocators, which sequentially allocate different sizes of objects in a continuous
\\range~\cite{DieHarder}. They maintain freelists for different size classes to assist fast allocations, and are also called freelist-based allocators. Representatives of these allocators include both the Windows and Linux allocators, as well as Hoard~\cite{Hoard}, whose design features very limited security countermeasures. Even worse, some implementations may directly conflict with the goal of security. For instance, they place metadata immediately prior to each object, and reutilize the first words of a freed object to store pointers used by their freelists~\cite{DieHarder}. These designs will significantly increase the attack surface, since attackers can easily overwrite freelist pointers or other metadata to initiate attacks. Further details are presented in Section~\ref{sec: allocators}.  

BIBOP-style (``Big Bag of Pages''~\cite{hanson1980}) allocators belong to the second type of allocators. They allocate several pages to serve as a ``bag'', where each bag will be used to hold heap objects of the same size. The metadata of heap objects, such as the size and availability information, is stored in a separate area. These allocators, such as \texttt{jemalloc}~\cite{jemalloc, jemalloc2}, Vam~\cite{Vam}, Cling~\cite{Akritidis:2010:CMA:1929820.1929836}, the OpenBSD allocator (which may be referred to simply as ``OpenBSD'' in the remainder of this paper)~\cite{openbsd}, TCMalloc~\cite{tcmalloc}, and DieHarder~\cite{DieHarder}, avoid corruption of the metadata through isolation mechanisms. To the best of our knowledge, all existing secure allocators utilize the BIBOP-style.

Existing secure allocators, such as OpenBSD~\cite{openbsd}, Cling~\cite{Akritidis:2010:CMA:1929820.1929836}, and DieHarder~\cite{DieHarder}, avoid the use of freelists for small objects. Instead, they maintain a bag-based bitmap to indicate the availability of all objects within the bag. Although the bitmap mechanism reduces the memory consumption associated with tracking the status of heap objects, using only one bit for each object, it may impose significant performance overhead. If allocators utilize randomized allocation, this may impose an even larger overhead. For instance, the OpenBSD allocator randomly chooses one possible object inside a bag, upon every allocation. However, if this object is not available, it will sequentially search for another available object inside the same bag. In the worst case, the number of checks performed to search the bag can be proportional to the number of objects inside the bag (see Section~\ref{sec: openbsd} for more details). Furthermore, both OpenBSD and DieHarder may introduce false sharing problems~\cite{Hoard}, since multiple threads are sharing the same heap. For these reasons, secure allocators are typically much slower than performance-oriented allocators, although Cling is an exception that only focuses on use-after-free problems. Based on our evaluation of open-source secure allocators, OpenBSD imposes 22\% performance overhead, and DieHarder results in over 36\% slower runtime, on average. 

This paper introduces \FG{}, a secure BIBOP-style allocator that overcomes the performance issues of existing secure allocators. \FG{} may not impose the same randomization as existing secure allocators, but runs at nearly the same speed as one representative performance-oriented allocator---the Linux allocator. 

First, \textbf{\FG{} designs a novel memory layout that combines the benefits of both BIBOP-style and sequential allocators}. \FG{} takes the approach of BIBOP-style allocators: each bag, consisting of multiple pages, will hold objects with the same size class, while the object metadata is placed in an area separate from the actual heap. This design helps prevent attacks caused by corrupted metadata. At the same time, \FG{} designs a ``sequential bag placement'' by employing the vast address space of 64-bit machines: \FG{} maps a huge chunk of memory initially, then divides it into multiple heaps. Each heap will be further divided into multiple subheaps, proportional to the number of threads, and bags with increasing size classes will be placed sequentially, starting from the minimum size class to the maximum size class. This layout enables constant-time metadata lookup.
If one bag inside the current heap is exhausted, \FG{} simply services new requests from the equivalent bag in the next available heap. The detailed design is shown in Figure~\ref{fig:allocator}. 
For the purposes of security, \FG{} also randomizes the following parameters: bag size, heap starting address, and metadata starting address, all of which increase the difficulty of attacks. Also, guard pages are randomly inserted throughout, in order to defend against buffer overflows and heap spraying. 

Second, \textbf{\FG{} adopts the freelist idea from per\-for\-mance-oriented allocators, and applies the shadow memory technique based on its novel layout}. \FG{} discards the bitmap and hashmap designs of existing secure allocators, as they are not suitable for performance. As described above, bitmaps may incur significant performance overhead, which could be proportional to the size of the bitmap. Instead, using freelists can guarantee constant-time memory allocations and deallocations. \FG{} further utilizes single-linked lists in order to prevent cycles within the list, which avoids the issue of double frees. It utilizes freelists to manage freed objects, but places the freelist pointers into segregated shadow memory, such that they cannot be easily corrupted. 

Third, \textbf{\FG{} greatly reduces the number of \texttt{mmap} calls required for allocating both the bags, and the metadata required for managing these chunks}. This design not only avoids the performance overhead caused by performing a large number of system calls, but also saves kernel resources in managing numerous small virtual memory regions. For the purposes of security, \FG{} selectively places internal guard pages within each bag, based on a user-specified budget. 

Additionally, \FG{} also fixes several implementation weaknesses of existing secure allocators.

\paragraph{Contribution}
\noindent In short, this paper makes the following contributions.

\begin{itemize}
\item \textbf{A Faster Secure Allocator}. We developed \FG{} to be a faster secure memory allocator. \FG{} was designed with a novel memory layout. In addition, \FG{} also applies the freelist and shadow memory techniques, and reduces the number of unnecessary system calls, in order to improve performance. \FG{} also fixes multiple issues associated with existing secure allocators, including possible false sharing problems, and provides better reporting of double and invalid frees.

\item \textbf{Extensive Analysis of Secure Allocators}. We have provided an extensive analysis of the performance and security issues of existing secure allocators, such as OpenBSD, DieHarder, and Cling. Some understanding is obtained directly through examination of their source code.

\item \textbf{Extensive Evaluation}. We have performed a large number of experiments to verify the performance, memory overhead, and effectiveness of \FG{}. Experimental results show that \FG{} imposes less than 2\% overhead when compared to the Linux allocator, while providing significantly better security. Furthermore, \FG{} considerably outperforms representative secure allocators, OpenBSD and DieHarder. 
 	
\end{itemize}

\paragraph{Outline}
The remainder of this paper is organized as follows. Section~\ref{sec:background} presents background on heap-related vulnerabilities. Section~\ref{sec: allocators} further examines the advantages and disadvantages of several representative allocators, which motivate our work. Based on our detailed analysis, we discuss the key ideas of \FG{} and its threat model in Section~\ref{sec:overview}. Then, Section~\ref{sec:implementation} provides the detailed implementation of \FG{}, while Section~\ref{sec:experimental} evaluates its performance, memory usage, and effectiveness. Next, Section~\ref{sec:discussion} discusses the limitations of \FG{}. Finally, Section~\ref{sec:relatedwork} lists relevant related work, and Section~\ref{sec:conclusion} concludes.

\section{Background}
\label{sec:background}

This section provides a background of relevant memory vulnerabilities, as well as an extensive analysis of existing allocators. Familiarity with these memory vulnerabilities helps to understand how \FG{} defeats them, while the analysis also helps to recognize the differences between \FG{} and these existing allocators. 

\subsection{Heap-related Memory Vulnerabilities}
\label{sec:vulnerabilities}

\subsubsection{Heap Over-reads}
A heap over-read occurs when a program overruns the boundary of an object, possibly reading adjacent memory that was not intended to be accessible. It includes heap under-reads, where memory locations prior to the target buffer are referenced. Heap over-reads can occur due to a lack of built-in bounds-checking on memory accesses, particularly for C/C++ programs. They can cause erratic program behavior, including memory access errors, incorrect results, or a crash. They can also lead to security problems, including information leakage and denial-of-service attacks.

\subsubsection{Heap Overflows} A heap overflow occurs when a program writes outside of the boundary of an allocated object. As with heap over-reads, throughout the remainder of this paper, heap overflows will also be used to refer to the related problem of corrupting memory immediately prior to the allocated object. Buffer overflows can cause security problems such as illegitimate privilege elevation, execution of arbitrary code, denial-of-service, and heap smashing. 

\subsubsection{Use-after-frees and Double-frees}

Use-after-free occurs whenever an application accesses a previously deallocated object. A recent study shows that use-after-free errors are the most severe vulnerabilities of the Chromium browser, in terms of both the number of occurrences, and the severity of security impacts~\cite{lee2015preventing}. Double-frees are considered to be a special case of use-after-free, and occur when an object has been freed twice. Depending on the design of the specific allocator, use-after-free may cause execution of arbitrary code, loss of integrity, and denial-of-service attacks.  
 
\subsubsection{Invalid frees} For invalid frees, applications invoke \texttt{free()} on a pointer that was not acquired using heap allocation functions, such as \texttt{malloc()}, \texttt{calloc()}, or \texttt{realloc()}. Invalid frees can cause the execution of arbitrary code, intentional modification of data, and denial-of-service attacks. 

\subsubsection{Other Heap Errors} Other heap-related security vulnerabilities exist, including: initialization errors, failure of return values, improper use of allocation functions, mismatched memory management routines (e.g., \texttt{malloc}/\texttt{delete}), and uninitialized reads, all of which can lead to exploitable vulnerabilities. We must note that \FG{} is not designed to handle these vulnerabilities.

\subsection{Existing Secure Allocators}
\label{sec: allocators}

As described in Section~\ref{sec:intro}, memory allocators can be classified into two major types: bump-pointer and BIBOP-style allocators. 

\begin{figure*}[!ht]
\begin{center}
\includegraphics[width=5in]{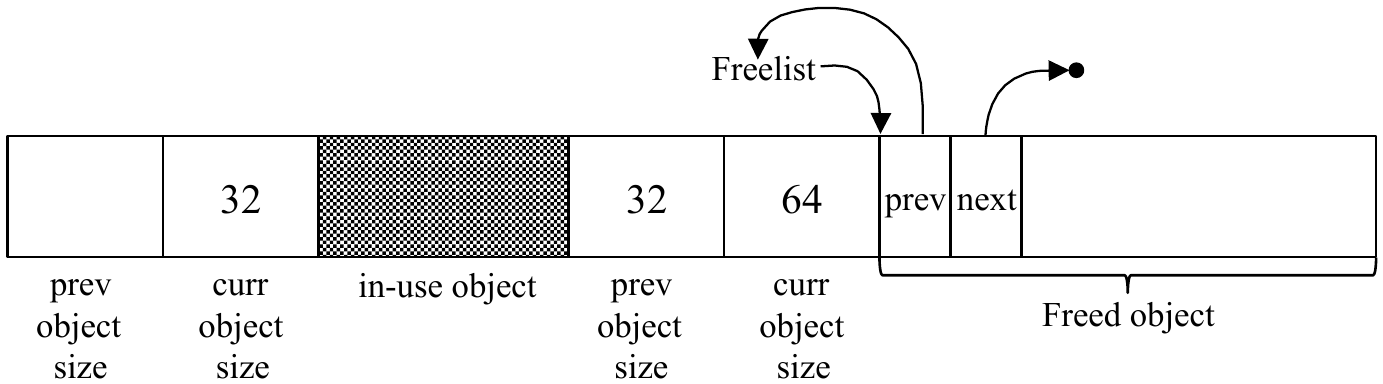}
\end{center}
\caption{
A fragment of the Linux allocator. Object headers are prepended to objects, which supports \\fast freeing and coalescing operations, but is vulnerable to overflows that can easily destroy the metadata. \label{fig:noramlallocator}}
\end{figure*}

\paragraph{Bump-pointer Allocators} Bump-pointer allocators, including the Windows and Linux allocators, typically employ freelists for the purpose of improved performance: they maintain freed objects in various freelists, organized by their size classes~\cite{DieHarder}. Figure~\ref{fig:noramlallocator} provides an overview of Linux's default memory allocator~\cite{dlmalloc}. However, as they were not designed for security, they actually increase the attack surface for malicious users. Metadata, such as size and status information, are prepended to heap objects, such that overflows can easily destroy their contents. To save space, they also embed freelist pointers directly within freed objects, which can be altered easily by buffer overflows and use-after-frees. The only security feature supported by the Linux allocator is the ability to detect double and invalid frees. However, even this feature is only partially achieved, as it checks a single bit embedded into the size field to confirm the status of the object. Furthermore, \DL{} has the following problems: (1) When the metadata is corrupted, due to buffer overflows or use-after-frees following consolidation, \DL{} may either miss problems or generate incorrect alarms. For example, it may report a normal free as an ``invalid free'' problem. (2) It cannot report invalid frees if the pointer is outside the range of valid heap addresses. (3) It may incorrectly report an invalid free as a double free problem if the pointer refers to an unallocated area.

\begin{figure*}[!ht]
\begin{center}
\includegraphics[width=5in]{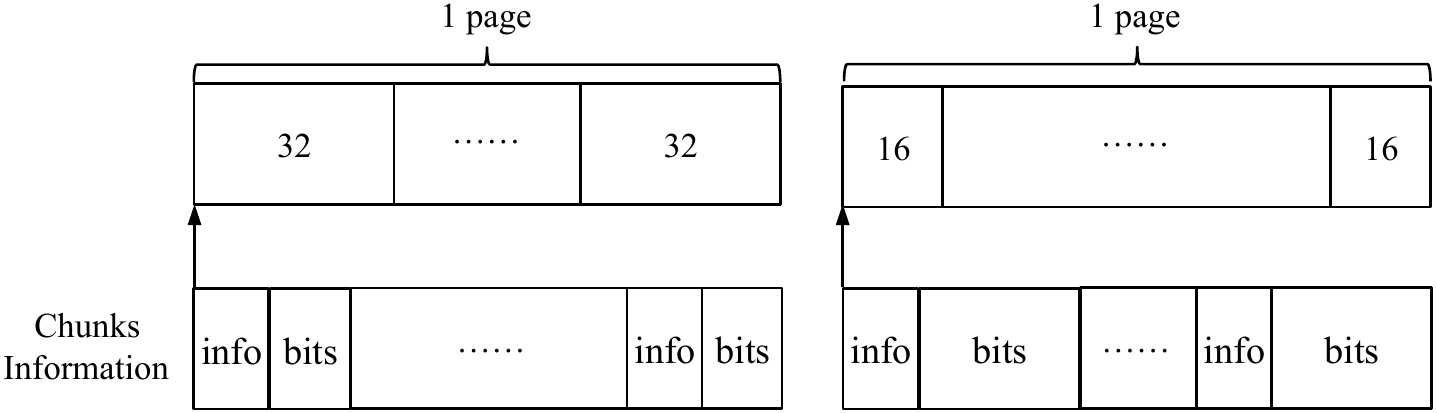}
\end{center}
\caption{
A fragment of the OpenBSD allocator. The mapping between bags  and metadata (e.g. chunks information) is kept in a global hash table, and each bag has a bitmap to maintain the status of all objects inside. Metadata are typically stored in a separate location. \label{fig:openbsd}}
\end{figure*}

\paragraph{BIBOP-style Allocators} BIBOP-style allocators, such as PHKmalloc~\cite{kamp1998malloc}, dnmalloc~\cite{dnmalloc}, Vam~\cite{Vam}, \texttt{jemalloc}~\cite{jemalloc, jemalloc2}, Open\-BSD~\cite{openbsd}, TCMalloc~\cite{tcmalloc}, Cling~\cite{Akritidis:2010:CMA:1929820.1929836}, and DieHarder~\cite{DieHarder}, typically allocate one or more pages at a time (known as a ``bag''), where each bag is used to hold heap objects of the same size. Among them, Cling, OpenBSD, and DieHarder are considered to be secure allocators, while others focus on performance only. We further discuss the design, advantages, and shortcomings of these three secure allocators.  
 
\subsubsection{OpenBSD Allocator}
\label{sec: openbsd}
The OpenBSD allocator originates from PHKmalloc~\cite{kamp1998malloc}, but features substantial improvements on the security~\cite{openbsd}. It avoids the use of freelists and inline metadata (headers). All descriptions here are based on the allocator of OpenBSD-6.0.

The OpenBSD allocator handles objects with small sizes differently than those with large sizes. Objects with sizes greater than 2 kilobytes will be considered as large objects. 

\paragraph{Management of Small Objects}
For small objects, the size of a bag is simply a page, which allows for multiple objects with the same size. For instance, for the 32-byte size class, 128 objects will fit inside one bag. OpenBSD allocates every bag by utilizing an \texttt{mmap} system call. It saves the information of chunks/objects in a separate area that is also obtained via the \texttt{mmap} system call. Typically, the information of multiple bags can share the same page, since the memory required to store the information for a single bag will be less than one page. As shown in Figure~\ref{fig:openbsd}, OpenBSD utilizes a bit in the bitmap (shown as ``bits'' in the figure) to indicate the status of every object, where $1$ indicates freed status, and $0$ represents in-use. Other bag information, such as the size and number of available objects, is stored in the area before the bitmap. The OpenBSD allocator utilizes a hash table to track the relationship between bags and their metadata information. Basically, given an address, we could obtain the starting address of the page, then use it to search the hash table to find the chunk information for this bag. The hash table will grow automatically, in order to reduce potential hash conflicts.

The allocation and deallocation of small objects are further described as follows. (1) During allocation, the OpenBSD allocator first randomly selects one-out-of-four bag lists for the given size class. If there are no available objects in the first bag, it will invoke \texttt{mmap} to first allocate another bag, then thread this bag into the bag list with the proper size class. When objects are available in the first bag, the allocator will select one object randomly from the bag, as discussed in ``Randomized Allocation'' below. (2) During deallocation, a freed object will be randomly placed into a delayed array that can hold up to 16 freed objects. If a previously-freed object already exists in that slot, the previously-freed object will be actually freed, and the newly-freed object will take its place in the delayed array. If the previously-freed object is the first freed object of its bag -- making the corresponding bag no longer full -- the bag will be moved into the header of this bag list. 

\paragraph{Management of Large Objects}
For large objects (whose sizes are greater than 2 KB), OpenBSD applies a different policy. By default, OpenBSD keeps at most 64 pages in the \texttt{free\_regions} cache in order to reduce the number of \texttt{mmap} system calls. Upon receiving an allocation request, OpenBSD will check whether it is possible to satisfy the request from the cached pages. If the requested size is less than the available pages in the cache, OpenBSD will check the entries in the cache, starting from a randomly-selected entry. If it can find one object whose size is equal to, or larger than, the requested size, OpenBSD allocates the object from the cache, and reinserts the remaining pages back into the cache. When there are no available objects capable of satisfying the current request from the cache, OpenBSD invokes the \texttt{mmap} system call directly.  

For deallocation, OpenBSD first checks whether the size of the freed object is larger than the size of the preset cache (64 pages). If so, then this object will be deallocated directly by invoking the \texttt{munmap} system call. Otherwise, the current object is added to a random location in the cache. Note, that if the current freed object increases the total size of freed objects in the cache beyond 64 pages, then some existing cached objects will be unmapped in order to limit the total size of freed objects in the cache to no more than 64 pages. 
 
Overall, the OpenBSD allocator implements the following approaches toward augmenting security, as shown in Table~\ref{tbl:comparison} as well. 

\paragraph{No freelist, no object headers, BIBOP-style} These properties are inherited from the original design of PHKmalloc~\cite{kamp1998malloc}. Since the OpenBSD allocator completely disposes with the use of freelists, it avoids possible corruptions to metadata (such as linked-list pointers) related to any freelists.

\paragraph{Fully-segregated metadata} Metadata information is maintained in an area separate from heap objects. This design is a departure from PHKmalloc, which stores metadata in the header of every chunk~\cite{kamp1998malloc}. Obviously, fully-segregated metadata helps augment security.

\paragraph{Sparse page layout/Guard pages} Rather than using \texttt{sbrk}, such as PHKmalloc, the OpenBSD allocator employs the \texttt{mmap} system call to allocate a page from the underlying operating system each time it is required for small objects. In effect, this mechanism effectively places unmapped ``guard pages'' between regions, which limits the exploitability of both overflow attacks and heap spraying attacks. 

\paragraph{Destroy-on-free} Destroy-on-free overwrites the contents of freed objects, filling them with random data. This policy is expected to locate some memory errors within applications. Currently, OpenBSD can also clean up an object prior to the memory being used, however, this feature is disabled by default due to performance considerations.

\paragraph{Randomized allocation} OpenBSD employs two types of randomization during allocation. Firstly, it maintains four lists for each size class, and chooses one randomly from among them. Secondly, inside a bag, it will determine the index of an allocation randomly. If the object with that index is in-use, it will search for the next available object, starting from the current position. Rather than employing a traditional linear search of each individual bit, it will sequentially test each of the bitmap's 16-bit \texttt{short} values until finding one whose value is non-zero. Then, it performs a fine-grained bit-level search of this \texttt{short} word to identify the next available object. The corresponding implementation is located between line 997 and line 1014 of \texttt{omalloc.c} of \texttt{OpenBSD-6.0}. Obviously, this second step may greatly compromise performance. In the worst case, when only a single free object exists in a bag, the number of searches is proportional to the total number of objects in the bag (e.g., a page). 

\paragraph{Delayed memory reuse} 
During memory deallocation, a freed object is placed into a delay buffer that can hold up to 16 objects with the same size class. It computes an index into this array randomly. If a previously-freed object is occupying the corresponding entry of the delay buffer, that object will actually be freed, making room for the currently-freed object to be placed into the delay array. 

\paragraph{Prevent invalid frees} The OpenBSD allocator could detect and prevent invalid frees with no false positives. Basically, it could identify the starting address and size of every object. If the corresponding object does not exist, or if the address is no longer a valid starting address, an invalid free is detected. Then, the allocator can stop execution for the purposes of attack prevention. 

\paragraph{Prevent double frees} The OpenBSD allocator has a \textbf{very low} probability of detecting double frees, due to an implementation issue. Currently, it only checks for a double free problem whenever a freed object is placed into the delay buffer. Only when the object in the selected slot of the delay buffer shares the same address as the newly-freed object will a double free problem be detected. However, it can tolerate double frees, since one bit is used to record the status of an object. It will not cause the same linked-list problem that is inherent in freelist-based allocators. 

\subsubsection{DieHarder}
\label{sec:dieharder}
DieHarder adapts many protections used by the OpenBSD allocator, but improves upon the randomized placement and randomized reuse by employing the randomization mechanism of DieHard~\cite{DieHard}. DieHarder sparsely utilizes the pages in a continuous range of virtual address space, which is different from DieHard. However, DieHarder does not place guard pages inside a continuous region. A buffer overflow cannot be detected, even if it may be tolerated by its over-provisioning mechanism.

To further tolerate the vulnerabilities imposed by buffer overflows, DieHarder guarantees that the ratio of allocated objects, to the number of total objects, will never exceed $1/M$, where $M$ represents the heap over-provisioning factor used to control this proportion. Thus, the entropy of choosing a random object is $\mathcal{O}(\log{}N)$ (where $N$ represents the number of allocated objects), which is much larger than that of the OpenBSD allocator. Similarly, DieHarder guarantees memory reuse to be less predictable. These two properties decrease the probability of overflow attacks. However, due to performance concerns, the default setting of $M$ is less than 2 (actually 8/7), which indicates DieHarder will not waste half of the heap space to achieve better security. 

DieHarder manages large objects differently from OpenBSD. Basically, it always allocates large objects using the \texttt{mmap} system call, then unmaps these objects by invoking the \texttt{munmap} system call. It does not utilize the cache mechanism of OpenBSD, thus helping  defeat use-after-free problems. However, it may impose much larger performance overhead, caused by numerous system calls. An object with size larger than 64 KB will be treated as a large object by DieHarder. Our experiments also confirmed that the size of large objects will have a significant impact on the performance of applications. 

DieHarder is capable of detecting double-frees and invalid-frees, but is currently configured to ignore them. It can tolerate these problems, which is the same as the OpenBSD allocator. DieHarder also has a scalability problem, as it uses a global lock to manage memory allocations/deallocations. This explains why the performance overhead reported here is higher than that of its original publication~\cite{DieHarder}, since all evaluated applications of Section~\ref{sec:performance} are multithreaded ones, instead of single-threaded applications, as in their paper. 

\subsubsection{Cling}
Cling is designed to defeat use-after-free problems, but does not protect against other types of vulnerabilities~\cite{Akritidis:2010:CMA:1929820.1929836}. It also utilizes the BIBOP-style for managing small objects, and its bitmap is located outside of the actual heap for security reasons. It borrows the type-safety memory allocation idea from existing work~\cite{Dhurjati:2003:MSW:780732.780743}, but without the use of compiler analysis. Basically, memory reuse is confined to only objects with the same type and same alignment (called ``confining memory reuse''). It avoids the use of freelists to mitigate the problem of metadata corruption. In order to avoid the scanning of bitmaps, Cling borrows the idea of ``reaps''~\cite{Berger:2002:RCM:582419.582421}, when multiple objects are allocated from the same allocation site. Basically, the allocations of objects inside the same bag will be in a sequential order. However, Cling does not introduce any randomization mechanism to increase the difficulty of other types of attacks. Also, the prevention of use-after-frees may fail if the object type is difficult to determine by allocation site~\cite{Undangle}.

\section{Overview}
\label{sec:overview}

\subsection{Key Ideas}

\label{sec:basicidea}

\begin{figure*}[!ht]
\begin{center}
\includegraphics[width=6.4in]{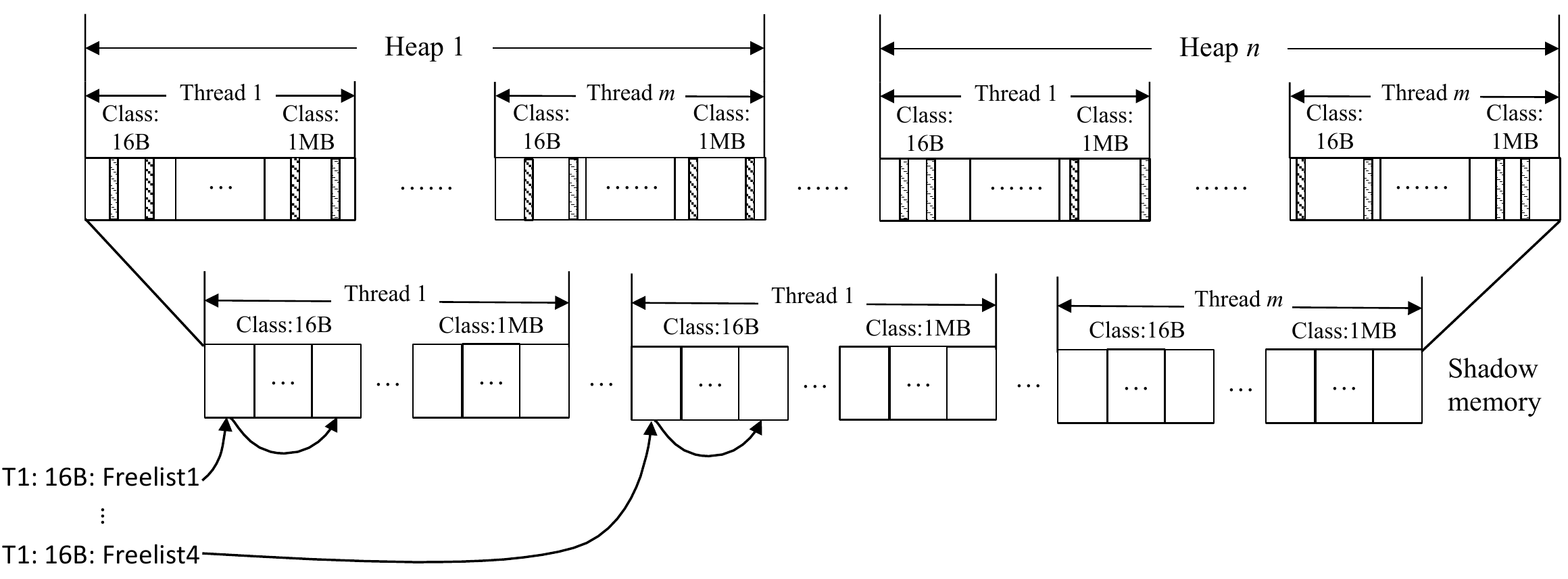}
\end{center}
\vspace{-0.2in}
\caption{
One example of \FG{}'s layout. \label{fig:allocator}}
\end{figure*}

To the best of our understanding, the OpenBSD allocator and Die\-Harder have the following issues or limitations:

\begin{itemize}
\item \textbf{Inefficiency caused by the use of bitmaps:} The bitmap design clearly reduces memory consumption, but compromises efficiency. For instance, the worst case when searching the bitmap for a free object is proportional to the number of objects inside a bag, due to their randomized allocation policy. 

\item \textbf{Reduced randomization for larger size classes:} Depending on the size class, the effective level of randomization in OpenBSD may not be uniform. A smaller size class will provide  better randomization, since an object will be chosen randomly from among the many objects within a page. However, when the size class is large, such as 2KB, only two objects are present in each page, and one will be selected randomly from one-out-of-four bags. 

\item \textbf{Inefficiency and extra memory overhead caused by \\page-based \texttt{mmap}:} The OpenBSD allocator invokes an \texttt{mmap} system call to allocate a chunk, as well as storage for its \texttt{chunk\_info}, every time. This method may place some guard pages between different chunks, due to the ASLR mechanism of the underlying operating system. However, a large number of system calls can significantly increase performance overhead. Also, the underlying OS may create a separate virtual memory region for each page, which consumes around 80 bytes of additional memory overhead in Linux. For small objects, DieHarder invokes \texttt{mmap} on the level of the miniheap, larger than one page, which reduces the number of system calls. However, it invokes \texttt{mmap} and \texttt{munmap} for large objects, whose size is larger than 64 kilobytes. In total, DieHarder also invokes a large number of \texttt{mmap} calls, as seen in Table~\ref{tbl:mmap}.

\item \textbf{False sharing performance problem:} In OpenBSD and DieHarder, multiple threads utilize the same heap simultaneously, which can cause false sharing problems that may substantially hurt the performance of applications~\cite{Hoard}. False sharing is a usage pattern in which two or more threads simultaneously access different objects co-located within the same cache line. If one of these threads modifies the data, this will result in the entire cache line being invalidated for the other threads, despite the fact that they were not using the exact data being modified. This cache invalidation will result in a costly re-fetch of the required data, degrading application performance~\cite{Sheriff}.

\item \textbf{Other problems:} DieHarder cannot detect double and invalid frees, based on our evaluation. The OpenBSD allocator can only report a very small portion of double frees, since it reports double-frees only when the object being inserted into the delay buffer shares the same address as the prior object occupying the same slot in the buffer. When configured to utilize canaries, the OpenBSD allocator will only check for overflow of freed objects, which is insufficient to stop many possible buffer overflow attacks. 

\end{itemize}

\paragraph{Our Design} Due to the above-mentioned problems, \FG{} designs a novel allocator aiming to balance performance and effectiveness. 

\FG{} adopts almost all security features listed in Table~\ref{tbl:comparison}, although with a lower entropy for randomization, as discussed in Section~\ref{sec:discussion}. The only feature not implemented by \FG{} is DieHarder's over-provisioned allocation. Over-provisioned allocation is useful to increase randomization and reduce attacks caused by buffer overflows, since overflows may occur in unallocated free space. However, over-provisioned allocation may significantly increase memory consumption, and largely decrease performance due to lower cache and memory utilization, combined with higher TLB pressure. Instead, \FG{} checks for the occurrence of overflow on neighboring objects at each deallocation, not just the item being freed, which is not supported by DieHarder. Then, if an overwrite is detected, \FG{} can stop the program immediately. This method helps thwart attacks caused by overflows in a more timely manner.

For performance reasons, \FG{} adapts the freelist mechanism that is widely utilized in performance-oriented allocators, such as the allocators of Linux and Windows systems. Freelists excel at performance, since each allocation and deallocation can be completed in constant time. Also, the freelist maintains the order of deallocations, which helps reduce attacks caused by use-after-frees, the most serious type of security attacks in Microsoft products recently~\cite{lee2015preventing}. Different from existing freelist allocators (see Section~\ref{sec: allocators}), \FG{} allocates these freelist pointers in a separate space, and uses only a single-linked list, to reduce memory consumption, shown as the shadow memory in Figure~\ref{fig:allocator}. To save space, object status information is stored within the same word: if the object is available, then its lowest-order bit will be 0 (this will hold true whether the location contains a pointer to the next available object, or whether it is null, indicating no next-available object exists). Conversely, if the object is in-use, its status will exactly equal 1. 

The \textbf{second design element} is to reduce performance overhead and memory consumption caused by page-based \texttt{mmap} operations. In order to reduce calls to \texttt{mmap}, \FG{} allocates a huge block initially, and places guard pages randomly inside each bag (shown as boxes with diagonal lines in Figure~\ref{fig:allocator}). Currently, guard pages will be placed randomly to occupy 10\% of each bag. This method reduces the number of \texttt{mmap} calls to less than 10\%, since OpenBSD invokes additional \texttt{mmap} system calls to allocate storage for \texttt{chunk\_info} structures, as well.

The \textbf{third design element} is to improve the performance of fetching corresponding metadata. Currently, OpenBSD and Die\-Harder create a hash table in which to map the page address of heap objects to a specific index, and grows the total size of this hash table whenever necessary. However, this still imposes significant performance overhead, especially when multiple pages are mapped to the same bucket. Instead, \FG{} relies on the fact that 64-bit machines have a vast address space, and utilizes the shadow memory technique to save metadata~\cite{qinzhao}. For any given heap address, \FG{} can quickly compute the location of its metadata, and vise versa.  The layout of the allocator is shown as Figure~\ref{fig:allocator}, and further described in Section~\ref{sec:implementation}.

\begin{table*}
  \footnotesize
  \centering
  \begin{tabular}{ l | l | c | c | c | c || c } 
  {Security Features} & \multicolumn{1}{|c|}{Security Properties} & {glibc} & Cling  & DieHarder & OpenBSD & \FG{} \\
   \hline
 
 No/segregated freelist  & {Prevent attacks on freelist related pointers}   &  & \checkmark & \checkmark  & \checkmark & \checkmark \\ \hline
 No object headers & {Prevent metadata related attacks} & & \checkmark & \checkmark & \checkmark & \checkmark\\ \hline
  BIBOP style & {Prevent metadata related attacks} & & \checkmark & \checkmark & \checkmark & \checkmark\\ \hline
  Fully-segregated metadata & {Prevent metadata related attacks} & & \checkmark & \checkmark & \checkmark & \checkmark\\ \hline
  Confining memory reuse & {Prevent use-after-free attacks} & & \checkmark &  &  & \\ \hline
 Destroy-on-free & {Help finding some memory errors} & &  & \checkmark & {\large $\diamond$} & {\large $\diamond$} \\ \hline
 \multirow{2}{*}{Guard pages} & {Reduce  attacks of buffer overflows and over-reads} & &  & \multirow{2}{*}{\checkmark} & \multirow{2}{*}{\checkmark}  & \multirow{2}{*}{\checkmark} \\ 
 & {Reduce heap spraying attacks} & & & & & \\ \hline
  Randomized allocation & {Increase difficulty of attacks caused by use-after-frees} & & & \checkmark & \checkmark  & \checkmark \\ \hline
  Over-provisioned allocation & {Reduce possible attacks caused by  overflows} & & & \checkmark & & \\ \hline
 Delayed/randomized reuse & {Reduce possible attacks caused by use-after-frees} & & & \checkmark &  $\ominus$ & \checkmark\\ \hline
 Detect invalid frees & {Prevent attacks caused by invalid frees} & $\ominus$ &  & \checkmark & \checkmark  & \checkmark \\ \hline
 Detect double frees & {Prevent attacks caused by double frees} & $\ominus$ & & \checkmark & $\ominus$  & \checkmark \\ \hline
Check overflows on frees & {Timely stop attacks caused by overflows} & & & & $\ominus$ & \checkmark \\ \hline
   \end{tabular}
  \caption{
  	Security features of existing secure allocators, with glibc added for comparison. ``\checkmark'' indicates the allocator has this feature. ``{\large $\diamond$}'' indicates this is an option, but is disabled by default.  ``$\ominus$'' indicates the implementation has some weakness. }
  \label{tbl:comparison}
\vspace{-0.2in}
\end{table*}

\subsection{Scope, Assumptions, and Threat Model}

The security properties supported by \FG{} are listed in Table~\ref{tbl:comparison}, as well as those of other allocators. Overall, \FG{} has the same performance overhead as the \texttt{glibc} allocator, but provides a better security guarantee than all existing allocators. Next, we discuss the attacks that can and cannot be stopped by \FG{}, and explain the fundamental reasoning.

\paragraph{Scope} For attacks based on invalid and double frees, \FG{} can prevent all such attacks, as long as the status of an object is never corrupted. Because the status information is kept in a separate location, this will greatly reduce the possibility of success for these attacks. Even if the status were to be modified by the attacker, some invalid frees caused by an invalid address can be prevented due to \FG{}'s special allocator design.  

Buffer overflow/over-read attacks will fail if the access touches one of the guard pages inserted randomly by \FG{}. Additionally, buffer overflows can be detected if one of the implanted canaries is found to have been corrupted. Implanting canaries will result in additional verification steps at the time the object (or one of its adjacent neighbors) is freed. At the same time, the difficulty of issuing these two types of attacks is increased due to randomized allocations, since the address of a target object is much harder to guess.

Attacks based on use-after-frees are reduced by utilizing delayed memory reuses. If an object is not re-utilized, the attacker may fail to exploit use-after-frees, since it will not cause any ill effect. Also, memory reuses are randomized to increase the difficulty of successful attacks.

\paragraph{Assumptions} \FG{} assumes that the starting addresses of both the heap and the shadow memory are kept hidden from the attacker. If an attacker has knowledge of these addresses, he can possibly change the status of an object, and force the allocator to make an incorrect decision. To avoid the predictability of these addresses, \FG{} allocates this memory using the \texttt{mmap} system call, which is guaranteed to return a random address if ASLR is enabled on the underlying OS. However, if the attacker has permission to run a program on the machine, he may be able to guess the location of the metadata, then take control of memory allocation. More discussion is provided in Section~\ref{sec:discussion}.

\section{Implementation}
\label{sec:implementation}

This section explains the detailed implementation of \FG{}. Basically, \FG{} focuses on the management of small objects, and adopts the same mechanism of DieHarder for managing larger objects. However, \FG{} defines large objects differently, such that only those objects with sizes larger than 1MB will be treated as ``large objects''. 

\subsection{Managing Small Objects}

\label{sec:smallobjects}

Section~\ref{sec:basicidea} describes the basic idea of managing small objects. First, \FG{} utilizes the BIBOP-style in order to place the metadata in another location, avoiding possible metadata-based attacks. This achieves the ``fully-segregated metadata'' target shown in Table~\ref{tbl:comparison}. Second, \FG{} utilizes freelists for better performance, rather than using a bitmap.  Third, \FG{} supports the fast fetching of metadata (such as freelist pointers) using a novel heap layout, shown as Figure~\ref{fig:allocator}.

\FG{} initially maps a huge block of memory, and divides this block into multiple heaps in the beginning. Inside each heap, \FG{} employs a per-thread subheap design so that memory allocations from different threads will be satisfied from different subheaps, in order to avoid possible false sharing problems~\cite{Hoard}. All bags belonging to a thread, which hold objects with different size classes, are located together. The bag size, starting address of the heap, and the starting address of the shadow memory that keeps the metadata of heap objects, are randomly chosen for each execution for the purpose of increased security. 

The rest of this section focuses on the implementation of other security features, as listed in Table~\ref{tbl:comparison}. 

\subsubsection{Randomized Guard Pages}
\FG{} initially utilizes the \texttt{mmap} system call to allocate a large chunk of memory, where the starting address of the heap is randomized between executions, a feature enabled by the ASLR mechanism of the underlying OS. The bag size utilized throughout each execution, which remains the same across the different size classes, is randomized with every execution, and ranges ranges between 4MB and 32MB. These mechanisms guarantee that the starting address of each bag is random across multiple executions.

\FG{} inserts guard pages randomly within each bag. Prior to allocating objects from a new page, \FG{} determines wheth\-er this page should be utilized as a guard page. This decision is based on a predetermined user budget, such as 10\%. Thus, 10\% of pages inside each bag will be chosen as guard pages. When a page is randomly selected to be a guard page, \FG{} invokes the \texttt{mprotect} system call to make this page inaccessible, such that all memory accesses on this page will be treated as invalid, and trigger segmentation faults. For a bag with a size class larger than one page (4KB), the size of its guard pages will be the same as the size class. That is, multiple pages will be utilized as guard pages in order to avoid misalignment of the metadata. Guard pages are useful for stopping buffer overflows, buffer over-reads, and heap spraying, as access on guard pages will immediately stop execution.

\subsubsection{Randomized Allocation and Delayed Reuse}

\FG{} takes a different approach from all existing allocators, by balancing randomization and performance. 

\FG{} maintains four bump pointers for each size class of each per-thread heap, which always point to the first never-allocated object~\cite{Lattner:2005:APA:1065010.1065027, Vam}. Objects will be allocated in a sequential order. After an object is allocated, the corresponding pointer will be bumped up to the next one. Whenever a bump pointer refers to the start of a new page, \FG{} determines wheth\-er this new page should be utilized as a guard page, as discussed above. \FG{} uses this sequential order for the purposes of performance, though it may compromise security. More discussion can be seen in Section~\ref{sec:discussion}. 

\FG{} also maintains four freelists to manage freed objects for each size class of each per-thread heap. A freed object will be added into one-out-of-four freelists randomly. Objects in a freelist will be reused in a first-in/first-out (FIFO) order. In this way, some use-after-free problems can be prevented automatically, since a freed object may be reallocated only after a long period, in which any use-after-free problems appearing in this period can be tolerated automatically. However, this method may slightly reduce performance compared with allocators using the last-in/first-out (LIFO) order. For the LIFO order, there is a significant chance that a newly allocated object is still inside the cache, which can avoid fetching from memory. However, our method will be superior to LIFO implementations in terms of security. It will significantly increase the difficulty of guessing the address of an allocation, due to the combination of FIFO and randomization, as discussed below. Overall, the FIFO mechanism increases both reliability and security. This mechanism cannot easily be supported when using bitmaps, such as the OpenBSD allocator or DieHarder. Bitmap-based allocators only use one bit to indicate the state of an object, either in-use or free. After a freed object is returned to the bitmap, there is no way to maintain the  temporal information. Due to the use of FIFO, there is no need to utilize a delay buffer, which is different from OpenBSD. 

\FG{} introduces randomization into its memory allocations. An allocation request could be satisfied either from one-of-four bump pointers, or one-of-four freelists, based on the value of a random number. This randomization is achieved through the following steps. First, we generate a random number $R$ using the Intel SSE2 number generator, as discussed below. We then take the modulus value $N$ by calculating $R \% 4$. $N$ will decide which freelist or bump pointer will be utilized. We will always check the $N$\textsuperscript{th} freelist first, and if freed objects are available, it will reuse them to satisfy the request. However, if there are no free objects in this freelist, the allocation will fall back to the $N$\textsuperscript{th} bump pointer. Furthermore, we will always check if the expression $R \% W$ is equal to zero, where $W$ represents a weighting factor. If so, \FG{} will strictly utilize the $N$\textsuperscript{th} bump pointer, regardless of whether the $N$\textsuperscript{th} freelist contains any objects available for reuse. Therefore, in terms of $W$, we will have a 1-in-$W$ chance of overriding the freelist and using the bump pointer instead. This method may slightly increase memory consumption and cause some slowdown, due to the increased memory footprint. However, it actually increases randomization, which is different from OpenBSD. OpenBSD will never allocate from a new bag, when there are freed objects it can reuse in the chosen bag. 

\paragraph{Incorporation of Fast Random Number Generator} 
In our initial design, we utilized the \texttt{glibc} \texttt{rand} function to generate a random number. However, this method is found to be very slow due to lock conflicts. The invocation of \texttt{rand} will acquire a global lock, which may prevent another thread from simultaneously obtaining a random number. To improve performance, \FG{} utilizes a \textbf{fast} pseudo-random number generator (RNG)\cite{sse2rng}. This faster RNG was optimized using Intel's SSE2 extensions, and further, does not require the use of synchronization primitives internally. Adopting this fast RNG reduced the performance overhead of \texttt{swaptions} by up to 65\%. 

\subsubsection{Checking Overflows at Deallocation}

\FG{} borrows another mechanism of OpenBSD to thwart possible attacks caused by buffer overflows. However, the OpenBSD allocator disables this mechanism, by default. In fact, based on our evaluation, this mechanism is very lightweight and helpful toward stopping attacks in a timely manner. 

\FG{} also increases the number of checks upon every deallocation. Currently, it will check the neighboring four objects as well, two before the current object and two after, instead of just one object. To support this, every allocation request will add one additional byte, at the end of the object, in which to hold a canary. Upon deallocation, if one of these five canaries has been changed to other values, \FG{} can halt execution of the current program. Note, that adding one byte to the end of an object may significantly increase memory consumption, since \FG{} always manages objects within size classes featuring powers of two. Thus, one additional byte may double the size of the memory consumption in the worst case. 

\subsubsection{Preventing Double and Invalid Frees}

For both of these problems, \FC{} will halt the execution immediately, and report the problem precisely, with 100\% guarantee. 

\FC{} prevents the following invalid frees: (1) If a free pointer lies outside the address range of the heap, a case which is easy to detect, and that most allocators can possibly detect. (2) If a free pointer falls within the range of the heap, but was never allocated. This could be discovered easily by checking its corresponding status. However, the Linux allocator may wrongly consider this problem to be a double-free error. \FC{} avoids this issue and reports it correctly. (3) If a free pointer is not aligned to the object's specific size class. \FC{} detects this problem easily based on its ``information computable'' design. \FG{} avoids false alarms and false negatives present in the Linux allocator, and caused by corruption of metadata, since \FC{} maintains the status of each object in shadow memory that is segregated from the actual heap. 

\FG{} also relies on the status information to detect possible double-frees upon deallocations. \FG{} always reports possible double frees, avoiding the implementation faults of the OpenBSD allocator. The segregation of metadata ensures that \FC{} can always detect double frees, unlike the Linux allocator.

\subsection{Managing Large Objects}

\label{sec:largeheap}
\FG{} borrows the same mechanism as DieHarder to handle large objects, which is discussed in Section~\ref{sec:dieharder}. Both provide better protection on ``large'' objects than OpenBSD. They can significantly reduce possible use-after-free attacks, since any access occurring after the \texttt{munmap} operation may actually cause the program to crash. They could defeat most buffer over-writes and over-reads, since the ASLR mechanism will effectively place guard pages before and after a mapped area, in most situations. Instead, OpenBSD maintains a cached list to track freed objects, which makes it fail to defeat use-after-frees. It has an option to protect the area of freed objects, but is disable by default due to performance reasons. We enabled this option and found that it may significantly affect performance, due to the increased number of system calls. OpenBSD treats objects with sizes larger than 2,048 bytes as large objects, resulting in many of its objects being treated as large objects. 

\FG{} defines ``large'' objects differently than DieHarder, which treats objects with sizes exceeding 64 kilobytes as large. This provides better protection than \FG{}, but with increased overhead due to the increased number of system calls.

\section{Experimental Evaluation}
\label{sec:experimental}

This section focuses on the following research questions. 
\begin{itemize}
\item What is the performance overhead of \FG{}, in comparison to the representative general-purpose allocator (the Linux allocator), and other secure allocators, such as the OpenBSD allocator and DieHarder?
\item What is the memory overhead of \FG{}? Also, we compare it against the allocators mentioned above.  
\item How effectively can \FG{} reduce or prevent real attacks? 
\end{itemize}

We performed all experiments on a 16-core quiescent machine, with two sockets installed with Intel(R) Xeon(R) CPU E5-2640 processors. This machine has 256GB of main memory, with 256KB L1, 2MB L2, and 20MB L3 cache. The experiments were performed using the unchanged Ubuntu 16.04, installed with Linux-4.4.25 kernel. We used GCC-4.9.1 with \texttt{-O2}, \texttt{-g} flags to compile all applications and all evaluated allocators appearing in this paper. 

We utilized the default settings for both the Linux allocator and DieHarder. For the OpenBSD allocator, we utilized a junking level of 0, in order to provide a fair comparison with \FG{}. For DieHarder, we utilized the version of 08/05/2017, where an object with size larger than 65,536 bytes (64 kilobytes) will be considered a large object, and with the heap multiplier $M$ set to $8/7$. We experienced much higher performance overhead when $M$ is set to 2, or higher. Both \FG{} and OpenBSD do not enable destroy-on-free, which is enabled by DieHarder.

\subsection{Performance Overhead}
\label{sec:performance}

\begin{figure*}[!th]
\begin{center}
\includegraphics[width=6.5in,height=3.2in]{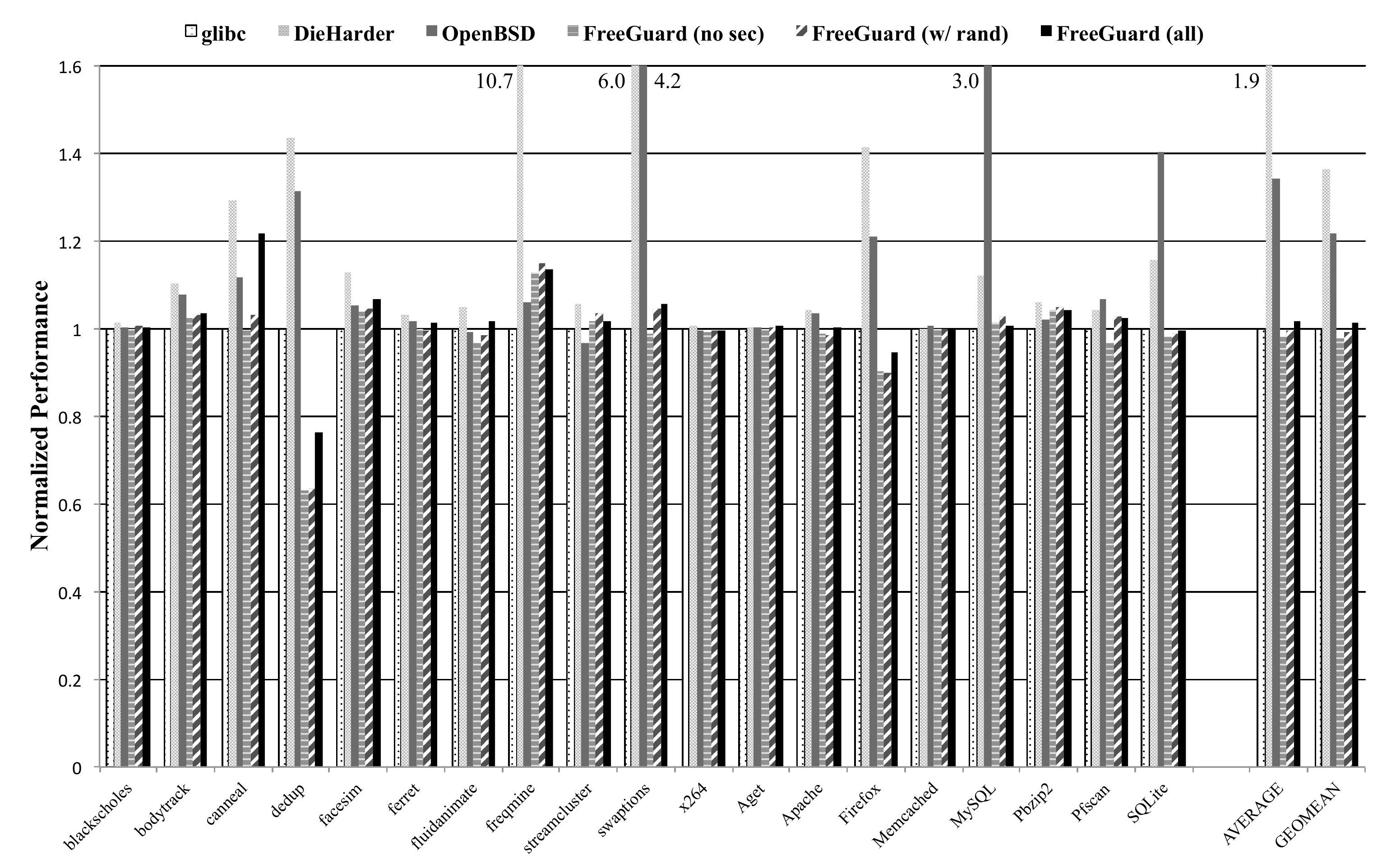}
\end{center}
\vspace{-0.2in}
\caption{
Normalized runtime with different allocators, where a higher bar indicates a higher overhead. \label{fig:performance}}
\end{figure*}

We evaluated 19 applications, and show the average results of ten executions in Figure~\ref{fig:performance}, where all values are normalized to the \texttt{glibc} library. A taller bar indicates a larger overhead. Among them, eleven are from the PARSEC suite of applications, while others are real applications, including \texttt{Apache httpd-2.4.25}, \texttt{Firefox-\allowbreak 52.0}, \texttt{MySQL-5.6.10}, \texttt{Memcached-1.4.25}, \texttt{SQLite-3.12.0}, \texttt{Aget}, \texttt{Pfscan}, and \texttt{Pbzip2}. All evaluated applications are multithreaded applications, making them more relevant toward gauging performance on modern multicore machines than single-threaded benchmark suites, such as SPEC. Both DieHarder and OpenBSD utilize a single heap to satisfy requests, instead of per-thread heaps, with a scalability issue. 

PARSEC applications were exercised using native inputs~\cite{parsec}. \texttt{MySQL} was tested using the \texttt{sysbench} application, with 16 threads and $100,000$ max requests, the throughput of which is shown. \texttt{Mem\-cached} was tested using the \texttt{python-memcached} script~\cite{memcached}, but changed to loop 20 times in order to obtain sufficient runtime. \texttt{SQLite} was tested using a program called ``threadtest3.c''~\cite{sqlitetest}. \texttt{Apache} was tested by sending $10,000$ requests via \texttt{ab}~\cite{apachetest}. For \texttt{Aget}, we collected the execution time of downloading 600MB of data from another quiescent server located on the local network. For \texttt{Pfscan}, we performed a keyword search within 800MB of data. For \texttt{Pbzip2}, we performed compression on a file containing 150MB of data. Finally, \texttt{Firefox-52.0} was evaluated in a headless configuration using a Python script to instruct the browser, via geckodriver, to fetch a fixed set of 71 web pages cached on a proxy server located on the local network. The \texttt{time} utility was used to measure the runtime required to perform these operations, as well as the maximum resident set size. Rather than utilizing \texttt{glibc}, Firefox uses its own default allocator based on \texttt{jemalloc}.

Figure~\ref{fig:performance} shows the normalized runtime of different allocators. Compared to the Linux allocator, \FC{}'s performance overhead is only 1.8\% using the arithmetic average, and 1.4\% using the geometric mean, with a number of security features enabled (Table~\ref{tbl:comparison}). In comparison, the OpenBSD allocator has an overhead of around 34\% (arithmetic mean) or 22\% (geometric mean), while DieHarder runs around 88\% (arithmetic mean) or 36\% (geometric mean) slower than the default Linux allocator. This indicates that \FG{} significantly outperforms the existing allocators. 

With the exception of \texttt{canneal} and \texttt{freqmine}, \FG{} imposes less than 10\% performance overhead. \FG{} has additionally enabled the delayed memory reuse feature by default, which adds around 2\% performance overhead.

As shown in Table~\ref{tbl:meminfo}, \texttt{canneal} involves a large number of memory allocations and deallocations, around 30 million. Thus, the performance slowdown is caused by the additional overhead of these allocations and deallocations. Since the glibc allocator prepends metadata before the actual objects, it takes virtually no time to fetch the metadata when there are no errors. Although \FG{} can compute the object's metadata location easily, it still imposes a larger overhead than the glibc allocator. For each deallocation, \FG{} must identify the placement of the metadata in order to add the entry into the freelist. When an object is allocated from a freelist, it must compute the corresponding heap address for the metadata. Currently, the freelist only contains the metadata address of the free objects. Afterwards, the metadata should be changed to reflect the object's updated status. The allocation of an object will actually involve two cache lines, instead of only one, which also adds some overhead.

For \texttt{freqmine}, as seen in Table~\ref{tbl:meminfo}, a considerable proportion of these allocations were large objects, which \FG{} handles by invoking the \texttt{mmap} system call in response to each such request. Therefore, a clear performance penalty will be associated with this method of handling large objects, and explains the degraded performance of \FG{} for this application.

Figure~\ref{fig:performance} also shows that \FC{} considerably outperforms the Linux allocator for the \texttt{dedup} application.  Based on our analysis, the Linux allocator's default configuration invokes a large number of \texttt{madvise} calls (over $500,000$), in order to return memory back to the OS. However, \FG{} does not invoke such \texttt{madvise} system calls, which explains why \FG{} significantly outperforms the Linux allocator in this case. Consequently, \FG{} shows larger memory consumption on this application, with around 64\% more memory used.

\subsection{Memory Overhead}
We have evaluated memory overhead of different allocators on the same applications. For server applications like \texttt{MySQL} and \texttt{Mem\-cached}, 
we executed a script to periodically collect the \texttt{/proc/\allowbreak PID/\allowbreak status} file, and utilize the maximum value of the VmHWM field to represent its maximum memory consumption. Memory consumption of other applications was collected using the \texttt{maxresident} output of the \texttt{time} utility, which reports the maximum amount of physical memory consumed by an application~\cite{timeutility}. 

\begin{table*}[tbp]
\centering
\begin{tabular}{l|c|r|rrr|rrrr}
\hline
\multirow{2}{*}{Programs} & Runtime  & Total Allocs & \multicolumn{3}{c|}{Large Allocations (\#)} &                \multicolumn{4}{c}{Memory Usage (MB)}        \\ \cline{4-10}
     & (s) &  \multicolumn{1}{c|}{(\#)} & \multicolumn{1}{c}{DieHarder} & \multicolumn{1}{c}{OpenBSD}   & \multicolumn{1}{c|}{\FG{}}          & \multicolumn{1}{c}{glibc}   &  \multicolumn{1}{c}{DieHarder} & \multicolumn{1}{c}{OpenBSD} &\multicolumn{1}{c}{\FG{}}  \\ \hline
blackscholes & 37.26 & 22 & 4 & 4 & 4 & 613 & 619 & 613 & 615 \\  
bodytrack & 26.6 & 437572 & 13053 & 15417 & 0 & 33 & 42 & 31 & 62 \\  
canneal & 55.39 & 30728188 & 1 & 1720 & 1 & 943 & 1131 & 808 & 1281 \\  
dedup & 16.23 & 4073908 & 359 & 1152213 & 7 & 1639 & 2076 & 1007 & 2830 \\  
facesim & 70.41 & 4746623 & 16970 & 33393 & 26 & 323 & 393 & 341 & 376 \\  
ferret & 4.58 & 139013 & 1557 & 7374 & 1 & 66 & 90 & 67 & 101 \\  
fluidanimate & 29.67 & 229928 & 6 & 8 & 2 & 408 & 464 & 429 & 433 \\  
freqmine & 44.31 & 6638 & 6103 & 6227 & 3068 & 1859 & 1785 & 1821 & 1996 \\  
streamcluster & 62.08 & 113271 & 30 & 1758 & 3 & 111 & 115 & 111 & 115 \\  
swaptions & 19.98 & 48001811 & 0 & 16000129 & 0 & 9 & 12 & 7 & 12 \\  
x264 & 53.23 & 35771 & 241 & 35483 & 35 & 485 & 516 & 502 & 483 \\ \hline \hline
Aget & 5.5 & 50 & 0 & 18 & 0 & 6 & 6 & 3 & 5 \\  
Apache & -- & 495 & 0 & 4 & 0 & 5 & 5 & 5 & 6 \\  
Firefox & 78.01 & 21126988 & 5393 & 413270 & 264 & 158 & 158 & 161 & 160 \\  
Memcached & 4.62 & 118 & 5 & 17 & 0 & 6 & 9 & 6 & 9 \\  
MySQL & -- & 1898867 & 51697 & 391172 & 5 & 123 & 132 & 271 & 154 \\  
Pbzip2 & 1.46 & 1229 & 851 & 1022 & 0 & 94 & 100 & 95 & 255 \\  
Pfscan & 1.46 & 43 & 0 & 2 & 0 & 735 & 782 & 784 & 821 \\  
SQLite & 20.62 & 1345226 & 9635 & 1075648 & 0 & 40 & 62 & 34 & 122 \\   \hline
\end{tabular}
\caption{Program characteristics related to different memory allocators.\\ (Apache and MySQL were measured by transactions per second rather than runtime.)\label{tbl:meminfo}
}
\vspace{-0.2in}
\end{table*}

The physical memory overhead of running Linux, OpenBSD, and \FC{}, is shown in Table~\ref{tbl:meminfo}. Overall, OpenBSD has almost the same memory overhead as Linux, since it always prefers freed objects. Comparing to the Linux allocator, the total memory overhead of \FG{} is around 20\%, while OpenBSD actually uses 6\% less memory, and DieHarder's memory overhead is about 11\%.

We have investigated to determine the cause of this. \FG{} utilizes four bump pointers and four freelists. Memory reuse randomization may significantly reduce the re-utilization of a particular object. For instance, consider an application performing eight allocations, each of size 1MB, where one allocation follows after another deallocation.  
 The Linux allocator will immediately re-utilize the freed object, which only increases the memory footprint by 1MB, in total. However, \FG{} may utilize up to 8MB of memory, since the randomization may choose to allocate only from bump pointers, or a chosen freelist may not have any available objects, causing it to fall back to the bump pointer again. Although, when there are already multiple objects available in the freelists, memory overhead will be not significantly increased. \FG{} compromises memory overhead in order to achieve better randomization.

Currently, \FC{} imposes more memory overhead than the OpenBSD allocator. There is a balance between memory overhead and performance overhead: (1) The OpenBSD allocator treats memory allocations larger than 2KB as large objects, which will be allocated utilizing \texttt{mmap} every time (see Table~\ref{tbl:mmap}). For each allocation larger than 2KB, but less than 1MB, the OpenBSD allocator will waste, at most, one page. Instead, \FG{} utilizes power-of-two size classes to manage objects less than 1MB. Thus, it is possible to have larger internal fragmentation, with an upper bound of 50\%. In its default setting, \FG{} adds one byte for the canary, which helps find possible overflows in a timely manner, and stop the program accordingly. This additional byte may waste almost 50\% of the allocated space if the original size was already aligned to a power of two. (2) The OpenBSD allocator utilizes one bit to indicate the status of an object, which also minimizes memory consumption, but with higher performance overhead. \FG{} will use one word for each object in order to thread an object into the freelist. (3) \FG{} utilizes four freelists and four bump pointers, and may randomly choose to allocate an object from bump pointers despite free object availability. This also adds some memory overhead, but provides better protection due to increased randomization.

\begin{table*}
\begin{center}
\setlength{\tabcolsep}{4pt}

\centering
\begin{tabular}{l|rrr|rrr|rrr|rrr} \hline
              & \multicolumn{3}{c|}{glibc} & \multicolumn{3}{c|}{DieHarder} & \multicolumn{3}{c|}{OpenBSD} & \multicolumn{3}{c}{FreeGuard} \\ \hline
              & mmap  & munmap & mprotect & mmap    & munmap  & mprotect  & mmap    & munmap & mprotect & mmap   & munmap   & mprotect  \\ \hline
blackscholes  & 36    & 17     & 24       & 126     & 17      & 32        & 59      & 17     & 35       & 38     & 5        & 11023      \\
bodytrack     & 6585  & 6555   & 125       & 20469   & 19600   & 33        & 19661   & 20466  & 36       & 6557   & 6526     & 11177     \\
canneal       & 52    & 24     & 42       & 215307  & 27      & 31        & 154856  & 152849 & 34       & 33     & 2        & 33763     \\
dedup         & 650   & 862     & 273989       & 258295    & 363      & 28        & 434648     & 499668      & 31       & 35     & 2        & 42631     \\
facesim       & 105    & 39     & 42       & 33106   & 16940   & 31        & 14062   & 13705  & 584      & 38     & 6       & 12249     \\
ferret        & 319    & 297     & 183       & 8119    & 1839      & 34        & 7037    & 8532   & 549       & 294     & 263        & 11483     \\
fluidanimate  & 43    & 19     & 28       & 20663   & 30      & 32        & 14929   & 14725  & 35       & 34     & 3        & 12532     \\
freqmine      & 212   & 164     & 167       & 6848    & 6322     & 49        & 6187     & 6035    & 52       & 3165    & 3124      & 11460     \\
streamcluster & 107   & 80     & 90       & 257     & 102      & 92        & 201     & 136    & 95       & 33     & 2        & 11042     \\
swaptions     & 53    & 25     & 220       & 992     & 20      & 32        & 365      & 14     & 35       & 32     & 1        & 87760      \\
x264          & 286   & 269    & 39       & 1424    & 503     & 32        & 1016     & 943    & 35       & 34     & 3        & 11319     \\ \hline \hline
Aget          & 54    & 27     & 44       & 99      & 14      & 32        & 87      & 14     & 35       & 33     & 2        & 11023     \\
Apache        & 239      & 32       & 141         & 417        & 34        & 140          & 295        & 30       & 143         & 225       &   32       & 10125          \\
Firefox       & 12248  & 8916    & 207834   & 70947   & 14545    & 209854    & 93006   & 143198  & 208699   & 11845   & 8685      & 202429    \\
Memcached     & 39    & 8      & 25       & 214     & 5       & 23        & 97      & 1      & 24       & 34     & 1        & 11030     \\
MySQL         & 154   & 33     & 326       & 17239    & 14449      & 62        & 22248    & 49876     & 65       & 67     & 17        & 12079     \\
Pbzip2        & 120    & 92     & 143       & 1114     & 1037      & 37        & 939     & 880     & 40       & 33     & 1        & 11088     \\
Pfscan        & 41    & 2      & 30       & 83      & 2       & 34        & 76      & 2      & 37       & 36     & 2        & 11025     \\
SQLite        & 65    & 33     & 4160       & 14152     & 9665      & 33        & 239746      & 254387     & 36       & 38     & 7        & 14994    \\ \hline
\end{tabular}
\caption{System call counts, including both the application and the allocator.}
\label{tbl:mmap}
\vspace{-0.2in}
\end{center}
\end{table*}

\subsection{Effectiveness}

\label{sec:effectiveness}

\begin{table}
  \footnotesize
  \centering
  \setlength{\tabcolsep}{4pt}
  \begin{tabular}{| c | c | c | c | c | } 
  \hline
  	Application & Vulnerability & Reference & Original & \FG{} \\
     \hline
  	bc-1.06 & Buffer Overflow & Bugbench~\cite{bugbench} & Crash & Mitigation \\ \hline
    ed-1.14.1 & Invalid Free & CVE-2017-5357 & Crash & Prevention \\ \hline
    gzip-1.2.4 & Buffer Overflow & Bugbench~\cite{bugbench} & Crash & Mitigation \\ \hline
    Heartbleed & Buffer Over-Read & CVE-2014-0160 & Data Leak & Mitigation \\ \hline
    
    Libtiff-4.0.1 & Buffer Overflow & CVE-2013-4243 & Crash & Mitigation \\  \hline

  	\multirow{3}{*}{PHP-5.3.6}
    & Use-After-Free & CVE-2016-6290 & Crash & Mitigation \\
    \cline{3-5}
  	& Use-After-Free & CVE-2016-3141 & Crash & Mitigation \\

    \cline{3-5}
  	& Double Free & CVE-2016-5772 & Crash & Prevention  \\

  \hline

   {polymorph-0.4.0} & Buffer Overflow & Bugbench~\cite{bugbench} & Crash & Mitigation \\ \hline

    Squid-2.3 & Buffer Overflow & CVE-2002-0068 & Crash & Prevention \\  \hline
  
  \end{tabular}
  \caption{
  	Verifying \FG{} on several vulnerabilities.
  \label{tbl:effectiveness}}
  \vspace{-0.3in}
\end{table}

To verify the effectiveness of \FG{}, we have tested it on several different real-world vulnerabilities, as shown in Table~\ref{tbl:effectiveness}. Note that these vulnerabilities have also been evaluated in other works, such as FreeSentry~\cite{younan2015freesentry}. 

We confirmed whether \FG{} can prevent or mitigate latent problems in these applications. ``Prevention'' indicates that \FG{} completely avoids the problem, such as with double or invalid frees. ``Mitigation'' indicates that the possibility of successful attack is reduced, although there is no full guarantee that such a problem will always be avoided.

All vulnerabilities were confirmed in the original applications prior to linking with the \FG{} library. All applications, except \texttt{OpenSSH}, resulted in program crash. These problems include use-after-free, double-free, and buffer overflow problems; OpenSSH experiences an information leak. 

\paragraph{bc-1.06} bc, an arbitrary precision numeric processing language, contains a heap buffer overflow. We obtained a buggy version of this program from BugBench, a C/C++ bug benchmark suite~\cite{bugbench}. Bad input can trigger the buffer overflow, and will normally result in a program crash. An array requiring 512 bytes is requested, and an object of size 1,024 bytes is returned by \FG{}. The bug causes bc to access one element beyond the boundary of the array. \FG{} prevents a crash from occurring, as the overflow occurs within the slack/unused portion of the object, but does not reach either the canary or guard page located at the end of the allocated space.

\paragraph{ed-1.14.1} ed has an invalid free problem that can cause the program to crash, since the developers changed a \texttt{malloc}'d buffer for a static one, but forgot to remove the corresponding free operation. \FG{} can always report and prevent this problem, and print the call stack of the invalid free inside. 

\paragraph{gzip-1.2.4} gzip, the GNU compression and decompression program, contains a stack overflow problem. We converted this problem into a heap overflow. When it occurs, it causes the program to crash, since adjacent metadata will be corrupted. \FG{}, however, avoids the crash, due to its segregated metadata. 

\paragraph{Heartbleed} \FG{}'s protection against buffer over-read was validated against the Heartbleed bug, which results in the leakage of up to 64KB of data occurring from the heap. During our evaluation, we observed that the OpenSSL library will request approximately 33KB to use for receiving the client heartbeat request. Due to the nature of a BIBOP-heap, this results in \FG{} fulfilling the request with a 64KB object, as this is the smallest available bag (whose sizes follow powers-of-two) capable of satisfying the request. The malicious heartbeat request contains a falsified payload length value, indicating the payload is 64KB long, when in fact, it is empty. The server then allocates a new buffer in which to construct the heartbeat reply, and proceeds to copy 64KB from the start of the payload region within the request buffer, the amount indicated by its falsified header value. However, because the request data is stored in a buffer of size 64KB, and the payload section is not located at the beginning of the object, this results in a buffer over-read occurring. Normally, this would result in the leakage of uninitialized heap data. But, with \FG{}'s random guard pages enabled, the buffer over-read can result in a program crash immediately upon accessing the random guard page (if present) placed at the end of the object. The proportion of random guard pages inserted onto the heap is configurable, and was set to 10\% for our evaluation. As such, the Heartbleed attack was prevented 1-in-10 times, resulting in a program crash.

\paragraph{Libtiff-4.0.1} To validate \FG{}'s protection against buffer overflow vulnerabilities, a heap-based buffer overflow found in gif2tiff, a GIF-to-TIFF image conversion tool found in the libtiff library, was tested. When supplied with a specially-crafted image, gif2tiff will attempt to process the file, resulting in a crash. An attacker might exploit this vulnerability, and could potentially execute arbitrary code under the privilege level of the account used to run the process.
We reproduce the exploit by supplying a crafted GIF image as input. After linking to \FG{}, the program avoids the crash, instead reporting, "illegal GIF block type".

\paragraph{PHP-5.3.6} For PHP, two use-after-free vulnerabilities are triggered by dedicated XML data. They would allow attackers to cause a denial-of-service attack by crashing the program. We apply \FG{} to PHP and rerun the vulnerable code. One program prints the correct data, while the other prints a null value. Although the output is not correct in these cases, \FG{} prevents the program crash and, therefore, successfully prevents the denial-of-service attack. \FG{} provides protection due to its delayed and randomized reuse mechanisms; by reutilizing freed objects in FIFO order, as well as randomly choosing an object to return, certain use-after-free errors will not result in the corruption of in-use object data.

We additionally tested \FG{} with PHP when experiencing a vulnerability caused by unserializing malicious XML data, an issue that results in program crash due to a double-free error. We use a malicious PHP script to reproduce this vulnerability. With \FG{}, the program reports the complete call stack upon the second free.

\paragraph{polymorph-0.4.0} Polymorph, a filename converter, has a stack buffer overflow problem that was modified to use the heap for the purposes of our evaluation. This overflow is actually very similar to that of \texttt{gzip}, which will change the metadata and cause the program to crash when using the \texttt{glibc} library. As with \texttt{gzip}, \FG{} avoids the crash due to its segregated metadata.

\paragraph{Squid-2.3} The affected version of Squid -- a caching Internet proxy -- contains a heap buffer overflow. Squid allocates memory from which to build a title URL string, however, it fails to account for the extra space needed to URL-escape the string. Thus, a buffer overflow occurs when the program attempts to escape the string, and writes the result to an inadequately-sized heap buffer. For this case, \FG{} will always detect the overflow before the program crashes, since \FG{} can always find out the corrupted canaries in single-threaded programs, although conceptually, \FG{} can probabilistically prevent buffer overflows.

\textbf{Conclusion:} In each of these instances, \FG{} allows the programs to either run normally with no ill effects, or immediately halts execution and reports the problem to the user. As described above, \FG{} can always detect double and invalid frees. \FG{} prevents or mitigates buffer overflows due to the following mechanisms: first, \FG{}'s metadata segregation prevents the corruption of metadata; second, due to \FG{}'s power-of-two object class sizes, it tolerates a certain level of corruption, and; third, when the canary within an adjacent object has been detected to be corrupt, \FG{} immediately produces a warning and calls \texttt{abort}. \FG{} also mitigates use-after-free problems due to its reuse of freed objects in FIFO order.

\section{Limitations}
\label{sec:discussion}

Both \FG{} and DieHarder utilize the same mechanism for the management of large objects, which is safer than that of OpenBSD. Currently, OpenBSD cannot effectively defend against use-after-free vulnerabilities due to its cache mechanism, since freed objects are not protected after their deallocations. Instead, accessing a freed object in \FG{} and DieHarder will cause an access violation, since a freed object will be unmapped directly. 

For small objects, \FG{} has some limitations in randomized allocation, randomized memory reuse, and freelist protection, which are discussed as follows. 

\paragraph{Randomized allocation} \FG{}'s randomized placement is not as strong as that of DieHarder and OpenBSD. DieHarder provides $\mathcal{O}(\log{}N)$ bits of entropy for its randomized placement~\cite{DieHarder}, where $N$ represents the number of freed objects for this size class in existing miniheaps. OpenBSD-6.0 first chooses one-out-of-four lists, then allocates an object randomly inside a bag. Currently, the size of a bag is just one page. Thus, a bag can hold 256 objects with the size 16 bytes, and 2 objects with the size 2,048 bytes. Thus, the entropy associated with OpenBSD is currently between 3 bits to 10 bits. \FG{} only has 2 bits of entropy, as it will choose one-out-of-four lists randomly. 

\paragraph{Randomized memory reuse} DieHarder has the same entropy as its randomized allocation, $\mathcal{O}(\log{}N)$ bits~\cite{DieHarder}. OpenBSD-6.0 has a delayed buffer with 16 entries, which will provide an entropy of 4 bits. After that, no  randomization exists: a freed object will be always placed back into the same bag; it will always allocate an object from the first bag, if any objects are available there. \FG{} does not use the delayed buffer, but will put a freed object into one-out-of-four lists randomly. \FG{} introduces another mechanism to increase the complexity of attack upon memory reuses: freed objects will be utilized in a FIFO order, and \FG{} may still allocate never-allocated objects, even when there are some freed objects in freelists. We cannot easily quantify this entropy, but it should not be worse than OpenBSD.  

\paragraph{Freelist protection} 
The biggest problem of \FG{} is that its freelists are not protected. Thus, if an attacker modifies them, they may take control of the heap allocator, and issue successful attacks afterwards. The relationship between \FG{}'s metadata and heap objects is computable, when the starting address of the metadata is known. If the attacker has permission to run a program on the target machine, he may be able to examine the contents of \texttt{/proc/PID/maps}, and identify the placement of the metadata. However, if the attacker is unable to run programs on the target machine, the randomization increases the complexity of such attacks. Both OpenBSD and DieHarder improve the protection of their metadata, since the relationship between heap objects and their metadata is actually stored in a hash map. OpenBSD dramatically increases the difficulty of attacks, since every bag and bag metadata have the same storage unit of one page. 

\section{Related Work}
\label{sec:relatedwork}
\subsection{Secure Heap Allocators}
There are other secure heap allocators, apart from those discussed in Section~\ref{sec:background}. However, most focus primarily on a particular type of security issue.

Some previous work has focused on securing only the object metadata. Robertson et al. proposed to prepend canaries and checksums for the metadata in order to detect possible overflows~\cite{Robertson:2003:RDH:1051937.1051947}. Younan et al. proposed to secure the metadata, utilizing a hash table that is placed in a separate location~\cite{Younan05securityof}. Heap Server places the metadata in the separate address space of another process for better protection~\cite{Kharbutli:2006:CEP:1168857.1168884}. The dnmalloc allocator allocates a separate chunk to hold the metadata, and utilizes a separate lookup table to map the chunks to their metadata, which is similar to OpenBSD and DieHarder~\cite{dnmalloc}. Although these works can prevent some metadata-related attacks, they cannot mitigate attacks toward the heap itself, such as use-after-free attacks.

Furthermore, some works aim to increase the non-determinism of memory allocations and reuses, including the changing of starting addresses~\cite{Bhatkar03addressobfuscation:, PaX}, and shuffling the reuse order of freed objects~\cite{Kharbutli:2006:CEP:1168857.1168884}. \FG{} adopts some of these techniques, but provides more protections than these works.

\subsection{Defending One Specific Type of Errors}
Many security hardening techniques maintain and lookup the metadata at runtime to defend against certain problems. These examples include the checking of bounds information for validating array references~\cite{overflow:WIT, overflow:Baggy}, the confirmation of type information to validate cast operations~\cite{Lee:2015:TCV:2831143.2831149}, and the collection of object pointer information to perform
garbage collection~\cite{Rafkind:2009:PGC:1542431.1542438}.

Iwahashi et al. proposed a Petri-net based signature that helps to understand and detect double-free vulnerabilities~\cite{Iwahashi:2008:TAG:1432478.1432489}. Undangle assigns a unique label to each object, and tracks the propagation of these labels, employing dynamic taint analysis. Upon deallocation, Undangle determines unsafe dangling pointers based on the lifetime of dangling pointers~\cite{Undangle}. FreeSentry protects against use-after-free vulnerabilities through compiler instrumentation~\cite{younan2015freesentry}. FreeSentry tracks pointers pointing to every object, then invalidates these pointers when an object is freed, which imposes around 25\% overhead on average. DangNULL prevents use-after-free and double-free vulnerabilities~\cite{lee2015preventing}. Similarly, DangNULL traces the relationship between pointers and objects, and nullifies those pointers when their pointing-to objects are freed. DangNULL also utilizes compiler instrumentation in order to collect the relationship between pointers and objects. Overall, DangNULL's performance overhead is between 22\% to 105\%.   

However, these countermeasures typically defend against only one type of approach. In contrast, \FG{} provides more comprehensive protection against a range of errors, with lower performance overhead. 

\subsection{Employing the Vast Address Space}
Archipelego~\cite{Lvin:2008:ATA:1346281.1346296} trades the address space for security and reliability by randomly placing objects in the vast address space. Thus, the probability of overflowing real data can be effectively reduced. Cling also utilizes the vast address space to tolerate use-after-free problems~\cite{Akritidis:2010:CMA:1929820.1929836}. 
\FG{} utilizes the vast address space to map heap objects to their metadata through the shadow memory technique, in order to achieve better performance.

\section{Conclusion}
\label{sec:conclusion}

This paper presents a novel secure heap allocator, \FG{}, which provides significantly better performance than existing secure allocators. \FG{} designs a novel memory layout, reduces a large number of \texttt{mmap} calls, and borrows the ``freelist'' idea from performance-oriented allocators. Overall, \FG{} imposes negligible performance overhead (less than 2\%) over the Linux allocator, but features a range of additional security features. In contrast, the OpenBSD allocator (the representative secure allocator), imposes around 22\% overhead on average, with the worst case running $3.9\times$ slower than \FG{}. Finally, we have released the source code of \FG{}, which is available at \url{https://github.com/UTSASRG/FreeGuard}. 

\section*{Acknowledgements}

We would like to thank our shepherd, Hamed Okhravi, and anonymous reviewers for their valuable suggestions and feedback, which significantly helped improve this paper. We are also thankful to Emery Berger for his invaluable comments, especially regarding DieHarder and OpenBSD, and suggestions on an early draft of this paper. The work is supported by UTSA, Google Faculty Award, and the National Science Foundation under Grants No. 1566154 and 1453011. It is also supported by an AFOSR grant, FA9550-14-1-0119. The opinions, findings, conclusions or recommendations expressed in this paper are those of the author(s) and do not necessarily reflect the views of the National Science Foundation.

\bibliographystyle{ACM-Reference-Format}
\bibliography{refs,ref,tongping}
\end{document}